\documentstyle[12pt]{article}
\begin{document}
\newcommand{\be}{\begin{equation}}
\newcommand{\ee}{\end{equation}}
\begin{titlepage}
\title{{\em COBE} constraints on Kaluza--Klein cosmologies}
\author{V. Faraoni, F. I. Cooperstock and J. M. Overduin
\\ \\{\small \it Department of Physics and Astronomy, University
of Victoria} \\
{\small \it P.O. Box 3055, Victoria, B.C. V8W 3P6 (Canada)}}
\date{}
\maketitle   \thispagestyle{empty}  \vspace*{1truecm}
\begin{abstract}
A class of Kaluza--Klein cosmologies recently
proposed is compared with observations of the cosmic
microwave background. Some models are ruled out, while others turn out to be
viable, with the number of extra dimensions being constrained by
the {\em COBE} and Tenerife experiments.
\end{abstract}
\vspace*{1truecm} \begin{center}  To appear in {\em Int. J. Mod. Phys. A}
\end{center}     \end{titlepage}   \clearpage

Kaluza--Klein (KK) theory \cite{CollinsMartinSquires} has been revived in
recent years in the forms of supergravity \cite{SUGRA},
superstring theories \cite{SUSTR}, and theories with an extended
gravitational sector. Some of the early ideas
of Kaluza and Klein, like the presence of compact extra dimensions
and the geometrical origin of scalar and gauge fields survive in these
theories. Many of the papers currently published in the context of KK
theories deal with cosmology.

In this Letter we discuss a class of KK cosmological
models proposed recently by Cho (\cite{Cho,ChoPRLett} and references therein)
and Cho and Yoon \cite{ChoYoon} (see also \cite{YoonBrill}). These
models have not been studied in
sufficient detail to decide if they are realistic. In particular, no
prediction was given which, in conjunction with observations, makes these
models falsifiable. Here we fill this gap for simple
versions of these KK cosmologies. These models are seen as a possible
alternative to ordinary inflation in \cite{Cho}, because they may solve
some of
the problems of the standard big bang model (including the horizon
problem). In this sense the term ``generalized inflation''is used in
\cite{Cho}. In the following we use the word inflation in the
conventional sense. More precisely, this term denotes a regime in
which the scale factor of the (4--dimensional) spacetime $a(t) $
satisfies $\ddot{a}>0$ (a dot denoting differentiation with
respect to the comoving time $t$).
Currently, a successful mechanism for the generation of density
perturbations is available only in inflationary models of the universe
\cite{needinfl}, and we discard the non--inflationary models in
\cite{Cho}--\cite{ChoYoon} as non--viable. Should a non--inflationary
mechanism for the generation of density perturbations
turn out to be viable, our procedure of ruling out some models on the
basis of their being non--inflationary would have to
be reexamined. However, until this happens, we adopt
the conventional idea that the generation of structures in the universe
requires inflation \cite{needinfl}.

The conclusions that we reach on the viability of the cosmologies of
\cite{Cho}--\cite{ChoYoon} depend on the details of these models and,
strictly speaking, cannot be generalized. However, the cosmologies in
\cite{Cho}--\cite{ChoYoon} are representative of many models that appeared
in recent literature on KK cosmology \cite{otherKK}.

As a starting point, we consider the models of \cite{Cho,ChoPRLett}. The
$(4+d)$--dimensional spacetime manifold
is assumed to be the product $ M\otimes K$, where $M$ is 4--dimensional and
$K$ is a $d$--dimensional submanifold ($d\geq 1$). The
$(4+d)$--dimensional metric is given by
\be  \label{1}
\left(\hat{g}_{AB}\right)=\left(
\begin{array}{cc}
\hat{g}_{\mu\nu}  & 0 \\
0 & \hat{\phi}_{ab}
\end{array} \right) \; ,
\ee
where\footnote{For ease
of comparison with the literature on inflation, we use units
in which the speed of light and the reduced Planck constant assume the value
of unity. The Planck mass is related to the Newton constant $G$ by
$m_{pl}=G^{-1/2}$. These units differ from those
used in refs.~\cite{Cho}--\cite{ChoYoon}. Apart from this, and other minor
differences in the symbols, our notations coincide with those of
\cite{Cho}--\cite{ChoYoon}.} $A,B,$~...~$=0,1,$~...,~$(4+d)$,
$\mu,\nu$,~...~$=0,1,2,3$, and $a,b,...
=4,5,$~...,~$(4+d)$. It is assumed that
$\hat{g}_{\mu\nu}$ (called the ``Jordan metric'' in \cite{ChoPRLett}) is
a Friedmann--Lemaitre--Robertson--Walker (FLRW) metric on $M$
and $\hat{\phi}_{ab}$ is a
Riemannian metric on $K$, with the extra dimensions being spacelike. For
simplicity it is assumed here that no
gauge field corresponding to ``off--diagonal'' terms in the metric (\ref{1})
is present and that the torsion tensor vanishes\footnote{The last
assumption will be relaxed in the following.}. These simplifications are
widely used in KK cosmology (e.g. \cite{Freund}). In addition, the
stress--energy tensor of matter in $(4+d)$ dimensions is assumed to vanish
identically, because we are interested in the density perturbations
originating during the inflationary era of the universe, which is
believed to be
dominated by a scalar field (of geometrical origin in KK theory). During
this era, the other forms of matter can be neglected. Following \cite{Cho},
we introduce the quantities $\phi$ and $\rho_{ab}$ defined by
\be
\phi =\left| \mbox{det}( \hat{\phi}_{ab}) \right| \; ,
\ee
\be
\hat{\phi}_{ab}=\phi^{1/d} \rho_{ab} \; ,
\ee
where $|$det$( \rho_{ab})|=1$.  The theory is essentially
vacuum general relativity in $(4+d)$
dimensions and is described by the Lagrangian density \cite{Cho}
\be
{\cal L}^{(4+d)}=-\, \frac{m_{pl}^2}{16 \pi} \left( \hat{R}+
\Lambda \right) \sqrt{-\mbox{det}(g_{AB})} \; ,
\ee
where $\hat{\nabla}_{\mu}$ and $\hat{R}$
are, respectively, the covariant derivative operator and the Ricci
curvature of the metric $\hat{g}_{AB}$, and
$\Lambda$ is the cosmological constant of the $(4+d)$--dimensional
spacetime.

Dimensional reduction and the conformal transformation\footnote{See
\cite{conformal} for a proof of the necessity
and uniqueness of the transformation (\ref{conftrans}),
(\ref{newsigma}). This
tranformation is widely used in many theories unifying gravity with the other
forces.}
\be   \label{conftrans}
g_{\mu\nu}=\sqrt{\phi} \, \hat{g}_{\mu\nu}
\ee
together with the redefinition of the scalar field
\be         \label{newsigma}
\sigma=\frac{1}{2} \left( \frac{d+2}{d} \right)^{1/2} \ln \phi
\ee
lead to the ``Einstein frame'', in which the theory is described by the
Lagrangian density \cite{Cho,ChoPRLett}
\be   \label{newlagrangian}
{\cal L}'=-\frac{m_{pl}^2}{16\pi}\left[ R+\frac{1}{2}\, \nabla_{\mu}\sigma
\nabla^{\mu} \sigma +V( \sigma ) +\lambda \left( \left|\mbox{det}( \rho_{ab})
\right|-1\right)  \right] \sqrt{-g} \; ,
\ee
where $\nabla_{\mu}$ and $R$ are, respectively, the covariant
derivative and the Ricci curvature of the
FLRW metric $g_{\mu\nu}$ (designated the
``Pauli metric'' in \cite{ChoPRLett}),
$g\equiv \mbox{det}\left( g_{\mu\nu} \right) $ and \cite{Cho,ChoPRLett}
\be  \label{Chopotential}
V( \sigma )=R_K \exp \left( -\alpha \sigma
\right) +\Lambda \exp \left(-\beta \sigma \right) \; ,
\ee
\be     \label{coefficient}
\alpha=\beta^{-1}=\left( \frac{d+2}{d}\right)^{1/2} \; .
\ee
Here $R_K$ is the Ricci curvature of the metric $\rho_{ab}$ on the
submanifold $K$ and $\lambda$ is a
Lagrange multiplier introduced to obtain the constraint equation
det$( \rho_{ab})=1$. For simplicity, we omit from the Lagrangian density
(\ref{newlagrangian}) the gauge fields included in
\cite{Cho}--\cite{ChoYoon}. This is
justified by the fact that in the proposed cosmologies, the universe is
dominated by the dilaton \cite{Cho}.

In \cite{Cho} the theory
described by the Lagrangian density (\ref{newlagrangian}) is interpreted as
a 4--dimensional relativistic cosmology, where the dominant material source is
the massless, minimally coupled, scalar field $
\sigma$ self--interacting via the potential (\ref{Chopotential}), and it is
suggested that this theory may describe an inflationary
universe. However, for this interpretation it is necessary to consider the
renormalized field
\be  \label{correctsigma}
\bar{\sigma}=\frac{\sigma}{\sqrt{16\pi G}}
\ee
instead of $\sigma$. This point is crucial for our results
because the arguments of the exponentials in the scalar field
potential are changed by this renormalization. In fact, a failure to
impose this renormalization would lead to incorrect conclusions
regarding the compatibility of these cosmologies with observations.

We also introduce the quantities
\be          \label{barquantities}
\bar{R}_K=\frac{R_K}{16\pi G}  \;\;\;\;\;\;\; , \;\;\;\;\;\;
\bar{\Lambda}=\frac{\Lambda}{16\pi G}  \;\;\;\;\;\;\; , \;\;\;\;\;\;
\bar{\lambda}=\frac{\lambda}{16\pi G}  \; .
\ee
The Lagrangian density of the cosmological theory with the correct coupling of
the scalar field is then given by
\be
{\cal L}'=-\left[ \frac{m_{pl}^2}{16\pi}\, R+\frac{1}{2}\, \nabla_{\mu}
\bar{\sigma}\nabla^{\mu} \bar{\sigma} +W( \bar{\sigma} ) +\bar{\lambda}
\left( \left|\mbox{det}( \rho_{ab})
\right|-1\right)  \right] \sqrt{-g} \; ,
\ee
where
\be    \label{newpotential}
W( \bar{\sigma} )=\bar{R}_K \exp \left( -\sqrt{16\pi}\, \alpha \,
\frac{\bar{\sigma}}{m_{pl}}
\right) +\bar{\Lambda}\exp \left(-\sqrt{16\pi}\, \beta \,
\frac{\bar{\sigma}}{m_{pl}}\right) \; .
\ee
It is to be noted that the cosmological
constant $\Lambda$ has been absorbed in the potential of the new
scalar field as a result of the conformal transformation and
hence there is no cosmological constant in the 4--dimensional spacetime.
The Friedmann equation
\be
\frac{\ddot{a}}{a}=-\frac{4\pi G}{3} \left( \rho+3P \right)
\ee
combined with the expressions for the energy density and pressure of the
scalar field
\be
\rho=\frac{\dot{\bar{\sigma}}^2}{2}+W \;\;\;\;\;\;\;\; ,
\;\; \;\; \;\; \;\; P=\frac{\dot{\bar{\sigma}}^2}{2}-W
\ee
gives
\be      \label{conditioninflation}
\frac{\ddot{a}}{a}=-\frac{8\pi G}{3} \left( \dot{\bar{\sigma}}^2 -W \right)
\; .
\ee
As is clear from eq.~(\ref{conditioninflation}), a necessary
condition for inflation is $W>0$. We now
consider various possible scenarios,
corresponding to different values of $R_K$, $\Lambda$ and $d$, which
act as parameters of the theory. The possibilities that $R_K$ is positive,
zero, or negative are considered in \cite{Cho}, and we will not discard any of
these {\em a priori}. Different choices are motivated as follows. A
compact internal
space $K$ corresponds to $R_K<0$ in these notations \cite{Cho}. For
example, the case in which $K$ is a $d$--dimensional sphere
has been considered in the literature \cite{SUGRA,positiveR}. Various
KK models in which the submanifold $K$ is Ricci--flat have also been
considered (e.g. \cite{RandjbarWetterich}), and two in particular are
worth mentioning. One occurs
if there is only one extra dimension ($d=1$); the other is the case in which
the compactification of the $K$ submanifold is performed by
the points identification $x^a \rightarrow x^a+2\pi$,
in the spirit of Kaluza's original paper.
We have the following results:

\noindent {\bf Case~1) $W=W( \bar{\sigma})$}
\begin{itemize}

\item {\bf Case~1a) $\bar{R}_K=0$; $\bar{\Lambda}\leq 0$:} $W\leq 0$
and inflation does not occur.

\item {\bf Case~1b): $\bar{R}_K=0$; $\bar{\Lambda}>0$:}
the potential (\ref{newpotential}) reduces to a
single exponential. This kind of potential has been studied in the context of
power--law inflation (\cite{AbbottWise,LucchinMatarrese}--see also
\cite{LiddleLyth}). To be specific, the scalar field potential
\be         \label{PLIpotential}
U( \bar{\sigma})=U_0 \exp \left( -\sqrt{\frac{16\pi}{p}}
\frac{\bar{\sigma}}{m_{pl}}\right) \; ,
\ee
where $V_0$ and $p$ are constants, is required in order to have a scale
factor $a(t) \propto t^p$ (which describes inflation if $p>1$)
in the FLRW metric $g_{\mu\nu}$. The power law
inflationary scenario can be solved exactly, and the density perturbations
are described by a power law spectrum with index $n-1$, where $n=1-2/p$
(see \cite{LiddleLyth} for a review).
By comparing eqs.~(\ref{newpotential}),~(\ref{coefficient}) in the
case $\bar{R}_K=0$ with eq.~(\ref{PLIpotential}), one obtains immediately
\be   \label{nd}
n=1-\frac{2d}{d+2} \leq \frac{1}{3} \; .
\ee
The 1$\sigma$ results of the {\em COBE} \cite{COBE} experiment provide us with
the limit $ n=1.1\pm 0.5 $. The combined statistical analysis of the
Tenerife and {\em COBE} experiments \cite{Hancocketal} shows that, taking
1$\sigma$ limits from both experiments, they are consistent if $n\geq 0.9$.
Therefore, the cosmological model under consideration is ruled out.

\item {\bf Case 1c) $\bar{R}_K<0$; $\bar{\Lambda} \leq 0$:} $W\leq 0$ and
inflation does not occur.

\item {\bf Case 1d) $\bar{R}_K<0$; $\bar{\Lambda} >0$:} this case can be
reduced to the double
exponential potential studied recently by Easther \cite{Easther}. The
corresponding inflation can mimic most of the usual models of
inflation. The parameters in \cite{Easther} assume the values
\be
{\cal A}=\bar{\Lambda} \;,\;\;{\cal B}=-\bar{R}_K \;,\;\;
\xi=\sqrt{2}\beta\;,\;\;m=\frac{d+2}{d} \;  .                      \ee
Easther studies the constraint imposed by the observations of microwave
background anisotropies and concludes that if $\xi \geq 0.5$ no values
of $m$ and $\xi$ produce a viable perturbation spectrum. In the present
case, $d\geq 1$ implies $\xi\geq \left( 2/3 \right)^{1/2}\simeq 0.82$.
Therefore this scenario is ruled out.

\item {\bf Case 1e) $\bar{R}_K>0$; $\bar{\Lambda}=0$:} we compare
eqs.~(\ref{newpotential}), (\ref{coefficient}) and (\ref{PLIpotential}) to
obtain $p=d/ (d+2) <1$, which does not correspond to an inflationary scenario.

\item {\bf Case 1f) $\bar{R}_K>0$; $\bar{\Lambda}>0$:} this situation is not
covered in\footnote{A double exponential potential is also considered in
\cite{StarkovichCooperstock}, but the range of the parameters
considered there does not include case~1f).}
\cite{Easther} (due to the restriction $m>1$ in that paper).
It is unclear if this model is inflationary, the
answer depending on the relative amplitudes of $\bar{R}_K$,
$\bar{\Lambda} $ and the
value of $d$. In any case, the
potential (\ref{newpotential}) is not a single exponential, which excludes
the power law inflation \cite{LucchinMatarrese,EllisMadsen} expected in KK
theories \cite{Olive}. A detailed study of models (unperturbed, or including
density perturbations) with both $\bar{R}_K$ and
$\bar{\Lambda}$ positive has not been given, and without a
specification of the values of $\bar{R}_K$, $\bar{\Lambda}$, $d$, the
comparison with observations appears problematic, to say the least.
In the absence of a clear prescription for a meaningful
inflationary model, the most general case $\bar{R}_K,\bar{\Lambda}>0$ will
not be discussed here. However, we can test the model in a meaningul
approximation.
A high value of $d$ ($d\geq 36$--40) has been advocated by many
authors in order to
generate a large amount of entropy in the 4--dimensional universe
\cite{highd}. By approximating eq.~(\ref{newpotential}) for $d>>1$,
we obtain the exponential potential $W_0 \exp \left( -\sqrt{16\pi}
\bar{\sigma}/m_{pl} \right)$, where $W_0=\bar{R}_K+\bar{\Lambda}$.
The potential has the power law form (\ref{PLIpotential}) with
$p=1$, which corresponds to a coasting universe (scale factor linear in time)
and is not inflationary. This conclusion applies to all models with
$\bar{R}_K+\bar{\Lambda}>0$ if the number of extra dimensions is high.

\item {\bf Case 1g) $\bar{R}_K>0$; $\bar{\Lambda}<0$:} the same conclusion
as in case~1f).

\end{itemize}

\noindent So far, no explicitly inflationary solutions consistent
with observations have been found. We consider now the cosmologies
introduced in
\cite{ChoYoon}. In these models the starting point is Einstein--Cartan
theory in $(4+d)$ dimensions, with a non--vanishing torsion tensor. The
effective theory in 4 dimensions
is similar to the models in \cite{Cho,ChoPRLett}.
The cosmologies in \cite{ChoYoon} include two scalar fields
$\sigma$ and $\varphi$. The previous remark on the interpretation of this
theory as 4--dimensional relativistic cosmology dominated by scalar fields
applies again: it is possible to consider the usual Einstein equations
and a cosmological scenario only after the scalar
fields are redefined according to eq.~(\ref{correctsigma}) and
to the analogous $\bar{\varphi}=\varphi/\sqrt{16\pi G}$. The
potential $V( \sigma,\varphi)$ in eq.~(5.2) of ref.~\cite{ChoYoon} has to
be corrected
accordingly. The particular cases given by eqs.~(5.6) and (5.7) in
\cite{ChoYoon} appear to be particularly interesting, since it is claimed that
they may give rise to inflation. In the former situation, the correct
potential is
\be  \label{ChoYoon1}
W( \bar{\sigma},0)=\frac{m_{pl}^2}{16\pi}\left[ -\, \frac{Ad}{4} \exp \left(
-\sqrt{\frac{16\pi (d+2)}{d}}\, \frac{\bar{\sigma}}{m_{pl}}\right)
-\Lambda \exp \left( -\sqrt{\frac{16\pi d}{d+2}}
\, \frac{\bar{\sigma}}{m_{pl}}\right) \right] \; ,
\ee
where the constant $A$ is defined in \cite{ChoYoon}. We can draw some
conclusions immediately.

\noindent {\bf Case2: $W=W( \bar{\sigma},0)$}
\begin{itemize}

\item {\bf Case~2a) $A>0$; $\Lambda \geq 0$:} $W<0$ and
there is no inflation.

\item {\bf Case~2b) $A>0$; $\Lambda <0$:} this corresponds again to the case
considered in \cite{Easther} with parameters
\be
{\cal A}=\bar{\Lambda} \;,\;\;{\cal B}=\frac{Ad}{4}\,
\frac{m_{pl}^2}{16\pi} \;,\;\;
\xi=\left( \frac{2d}{d+2}\right)^{1/2}\geq 0.82\;,\;\;m=\frac{d+2}{d} \;.
\ee
The situation is the same as in case~1d) and this scenario is excluded as well
by the microwave background experiments.

\item {\bf Case 2c) $A=0$; $\Lambda \geq 0$:} $W\leq 0$ and inflation
does not occur.

\item {\bf Case 2d) $A=0$;
$\Lambda <0$:} the potential reduces to the form
(\ref{PLIpotential}) and there is power law inflation with $p=(d+2)/d$. This
scenario is ruled out by {\em COBE} analogously to case~1b).

\item {\bf Case~2e) $A<0$;
$\Lambda =0$:} the potential reduces to the form
(\ref{PLIpotential}) with $p=d/(d+2)$, which is not inflationary.

\item {\bf Case~2f) $A<0$;
$ \Lambda \neq 0$:} the same conclusion as in case~1f), but now
$W_0= \left( 64\pi \right)^{-1} \left| A\right| d m_{pl}^2-\bar{\Lambda}$.

\end{itemize}

\noindent Again, no explicitly inflationary universes consistent
with observations have been obtained in case~2).

 If, instead, the potential given in eq.~(5.7) of
\cite{ChoYoon} (corrected
after the rescaling of $\varphi$) is adopted, one has
\begin{eqnarray}  \label{ChoYoon2}
W(0,\bar{\varphi})=\frac{m_{pl}^2}{16\pi}\left\{ \frac{A}{4}\left[ \exp
\left(-\sqrt{\frac{32\pi}{d(d-1)}}\,(d+1) \,\frac{\bar{\varphi}}{m_{pl}}
\right) \right. \right.  \nonumber \\
\left. \left. -(d+1) \exp \left( -\sqrt{\frac{32\pi}{d(d-1)}}
\, \frac{\bar{\varphi}}{m_{pl}}
\right) \right]-\Lambda \right\}   \; .
\end{eqnarray}
In this case the effective cosmological constant
$\Lambda_4=-\bar{\Lambda}$ is induced in the
4--dimensional spacetime.

\noindent {\bf Case 3: $W=W(0,\bar{\varphi})$}
\begin{itemize}

\item {\bf Case~3a)
$\Lambda =0$; $A=0$:} $W=0$ and there is no
inflation.

\item {\bf Case~3b) $W(0,\bar{\varphi})$;
$\Lambda =0$; $A>0$:} the same conclusion holds here as in case~1f), but
no conclusion can be reached for $d>>1$.

\item {\bf Case~3c) $W(0,\bar{\varphi})$;
$\Lambda =0$; $A<0$:} this case is reduced to the scenario
studied in \cite{Easther} with the values of the parameters
\be   \label{Eastherparameters}
{\cal A}=|A| (d+1) \, \frac{m_{pl}^2}{64\pi} \;,\;\;{\cal B}=|A|\,
\frac{m_{pl}^2}{64\pi}
\;,\;\; \xi=\frac{2}{\sqrt{d(d-1)}} \;,\;\;m=d+1 \;.
\ee
Easther \cite{Easther} concludes that if $m\xi^2 \leq 0.15$
(with $m>1$) the resulting
spectrum satisfies the {\em COBE} and {\em QDOT} (with cold dark matter)
constraints for all values of $\bar{\varphi}$. On the contrary,
the perturbation spectrum is incompatible with these observations
if $\xi \geq 0.5$. The former inequality is equivalent to $d\geq 29$, and
the latter is equivalent to $ d \leq 4$. Therefore the model is not
viable if $d\leq 4$ and is viable if $d\geq 29$.

The analysis of the cases $d=5$,~...~,~28 requires the expression of the
spectral index of density perturbations in this potential, as computed
in \cite{Easther}:
\be \label{Eastherspectralindex}
n=1+\frac{m\xi^2}{\left( m-1+y \right)^2}\left[ 2(m-1)^2(y-1)-my^2
\right] \; ,
\ee
where the variable $y$ is related to the scalar field by
\be
1-y=\exp \left( -\sqrt{\frac{32\pi d}{d-1}} \,
\frac{\bar{\varphi}}{m_{pl}} \right)  \; .
\ee
In the relevant range of $y\in$[0,~1) \cite{Easther}, $n$ is a monotonically
increasing function of $d$, for all $d\geq 4$. The constraint $n\geq 0.9$
\cite{Hancocketal} is satisfied if $y\geq y_0$. The root $y_0$ of the
equation $n(y)=0.9$ is less than unity only if $d\geq 7$ and
$\bar{\varphi}\geq
\bar{\varphi}_0$, where $\bar{\varphi}_0$ is a decreasing function of $d$.
Therefore, the model is viable only if $d\geq 7$, and only over a range of
values of the scalar field $\bar{\varphi}$.
To summarize, this model cannot be viable for $d<7$. It is viable (for some
values of the scalar field $\bar{\varphi}$) over the range $7\leq d \leq 29$
and, for $d\geq 29$, it is viable in all cases.

It is remarkable that observations of the microwave
background constrain the number of extra dimensions in a higher dimensional
theory. This possibility is unique, since higher dimensional theories
\cite{SUGRA,SUSTR} are usually so complicated that the dimension cannot be
varied as a parameter. It is hoped that the KK models considered here serve as
a toy model for more fundamental theories, and that the restrictions on the
dimensionality of spacetime has a broader range of validity.
Cosmological perturbations in KK theories have been
considered in the literature \cite{KKperturbations}, but the possibility of
relating the number of extra dimensions to observable quantities has
apparently gone unnoticed until now.

\item {\bf Case~3d) $\Lambda >0$; $A=0$:} $W<0$ and there is no
inflation.

\item {\bf Case~3e) $\Lambda \neq 0$; $A\neq 0$:} the same as case~1f).
However, in the approximation $d>>1$ the potential (\ref{ChoYoon2}) reduces
to
\be
W(0,\bar{\varphi}) \approx \frac{m_{pl}^2}{16\pi}
\left[ \frac{A}{4} \exp \left( -\sqrt{32\pi} \,
\frac{\bar{\varphi}}{m_{pl}}\right)-\left( \Lambda+\frac{Ad}{4}
\right) \right]  \; .
\ee
The cosmological constant $\Lambda_4=-\left[ (64\pi )^{-1}Ad m_{pl}^2+
\bar{\Lambda}\right] $ is induced in the 4--dimensional spacetime.
If $\Lambda =-Ad/4$, there is no cosmological constant in the
4--dimensional spacetime. Then, if $A<0$, $W<0$ and there is no inflation.
If instead $A>0$, the scale factor has the form $a(t) \propto \sqrt{t}$
describing a radiation--dominated universe.

\item {\bf Case~3f) $\Lambda <0$; $A=0$:} the potential corresponds to a
de~Sitter expansion.

\end{itemize}

\noindent Our results can be summarized as follows. We adopt the point of view
that inflation is a necessary ingredient for a successful cosmological
model. Then the cosmological
models of cases 1a), 1c), 1e) 1f) and 1g) for $d>>1$, 2e), 2c), 2e),
3a),
3d), 3e) if $d>>1$ and $\Lambda_4=0$ are ruled out because inflation
does not occur. The models of cases 1b), 1d), 2b), 2d), and 3c) if
$d\leq 6$ are
incompatible with observations of the cosmic microwave background
anisotropies. In general, the cases 1f), 1g), 2f), 3b) and 3e) require a more
detailed study that is beyond the focus of this work. The same
applies to the model of \cite{YoonBrill}, which is inflationary.
Case 3c) is viable if $d\geq 7$.
In this case, the {\em COBE} and Tenerife experiments
constrain the number of extra dimensions.

It is to be noted that if the renormalization (\ref{correctsigma}),
(\ref{barquantities}) is missed, and the incorrect potential given by
eqs.~(\ref{Chopotential}), (\ref{coefficient}) is used, one would obtain the
erroneous result that many of the models in {\cite{Cho} give rise to power law
inflation, with spectral indices of perturbations compatible with the {\em
COBE} and Tenerife experiments.

\section*{Acknowledgment}

This research was supported, in part, by a grant from the Natural Sciences
and Engineering Research Council of Canada.

\end{document}